\documentclass{article}

\usepackage{cite}
\usepackage{graphicx}
\usepackage{dcolumn}

\begin{document}

\date{}
\title{On two perturbation series for short-range attractive potentials}
\author{Paolo Amore \\
Facultad de Ciencias, CUICBAS,\\
Universidad de Colima, Bernal D\'{\i}az del Castillo 340, \\
Colima, Colima, Mexico\\
and Francisco M. Fern\'andez \\
INIFTA (CONICET, UNLP), Divisi\'{o}n Qu\'{i}mica Te\'{o}rica, \\
Blvd. 113 y 64 (S/N), Sucursal 4, Casilla de Correo 16, \\
1900 La Plata, Argentina}
\maketitle

\begin{abstract}
We compare two alternative expansions for finite attractive wells. One of
them is known from long ago and is given in terms of powers of the strength
parameter. The other one is based on the solution of the equations of the
Rayleigh-Schr\"{o}dinger perturbation theory in a basis set of functions of
period $L$. The analysis of exactly solvable models shows that although the
exact solution of the problem with periodic boundary conditions yields the
correct result when $L\rightarrow \infty$ the coefficients of the series for
this same problem blow up and fail to produce the correct asymptotic
expansion.
\end{abstract}

\section{Introduction}

\label{sec:intro}

In a recent paper Lisowski et al\cite{LNPSSG16} proposed the application of
an approximate method for the treatment of the Schr\"{o}dinger equation with
finite attractive potentials. It consists of solving the secular equation
for the matrix representation of the Hamiltonian operator in a basis set of
functions of period $L$. The eigenvalues of this matrix are expected to
approach the actual eigenvalues of the problem in the limit $L\rightarrow
\infty $. The authors also applied the same approach to the equations given
by perturbation theory thus obtaining approximate perturbation coefficients
that depend on the box length $L$. They argued that this perturbation series
is convergent when the width of the ground-state eigenfunction is larger
than $L$ and divergent when that width is smaller than $L$. The former case
takes place when the potential strength $V_{0}>0$ is smaller than a critical
value $V_{crit}$ and the latter when $V_{0}>V_{crit}$. In order to sum the
perturbation series in both regimes the authors proposed the application of
the well known Borel transformation with the substitution of a finite
integral for the infinite one when carrying out the inverse transformation.

It is well known that there exists a perturbation series about $V_{0}=0$ in
the case of a short-range potential and there are even explicit expressions
for the first perturbation coefficients\cite{P80} (and references therein).
However, Lisowski et al\cite{LNPSSG16} seem to believe that any perturbation
series is ``inapplicable to predict the energies for attractive
potentials''. The purpose of this paper is to discuss the connection between
the approximate perturbation series proposed by Lisowski et al\cite{LNPSSG16}
and the well known exact perturbation series\cite{P80}.

In section~\ref{sec:gen_wells} we outline the application of the two
perturbation methods just mentioned to a general finite attractive well. In
section~\ref{sec:examples} we discuss the perturbation series by means of
simple, exactly solvable models. Finally, in section~\ref{sec:conclusions}
we summarize the main results and draw conclusions.

\section{Short-range shallow wells}

\label{sec:gen_wells}

Throughout this paper we consider a particle of mass $m$ that moves in one
dimension under the effect of a short-range negative potential $-V_{0}\leq
V(x)\leq 0$ that we suppose to be of even parity $V(-x)=V(x)$. In order to
simplify the calculations it is convenient to rewrite the Hamiltonian
operator
\begin{equation}
H=-\frac{\hbar ^{2}}{2m}\frac{d^{2}}{dx^{2}}+V(x),
\end{equation}
in dimensionless form by means of the change of variables $x=\gamma
x^{\prime }$, where $\gamma $ is a suitable length. In this way we obtain
\begin{eqnarray}
H^{\prime } &=&\frac{2m\gamma ^{2}}{\hbar ^{2}}H=-\frac{d^{2}}{dx^{\prime 2}}%
+\lambda v(x^{\prime }),  \nonumber \\
\lambda &=&\frac{2m\gamma ^{2}V_{0}}{\hbar ^{2}},\;v(x^{\prime })=\frac{%
V(\gamma x^{\prime })}{V_{0}}.
\end{eqnarray}
Thus an eigenvalue $E$ of $H$ and the corresponding one $\epsilon $ of $%
H^{\prime }$ are related by
\begin{equation}
\epsilon =\frac{2m\gamma ^{2}}{\hbar ^{2}}E.
\end{equation}
From now on we omit the prime on the dimensionless quantities and write the
Hamiltonian operator as
\begin{equation}
H=-\frac{d^{2}}{dx^{2}}+\lambda v(x),  \label{eq:H_dim}
\end{equation}
where $-1\leq v(x)\leq 0$.

It is well known that in the case of a short-range potential the ground
state energy can be expanded in a formal perturbation series of the form\cite
{P80}:
\begin{eqnarray}
-(-\epsilon )^{1/2} &=&\frac{\lambda }{2}\int v(x)\,dx+\frac{\lambda ^{2}}{4}%
\int \int v(x)v(y)|x-y|\,dxdy+  \nonumber \\
&&+\frac{\lambda ^{3}}{48}\int \int \int v(x)v(y)v(z)\left(
|x-y|+|y-z|+|z-x|\right) ^{2}dxdydz  \nonumber \\
&&+\frac{\lambda ^{4}}{96}\int \int \int \int v(x)v(y)v(z)v(t)\left(
|x-y|^{3}\right.  \nonumber \\
&&+\left. 6|x-y|^{2}|x-z|+3|x-y|^{2}|z-t|\right.  \nonumber \\
&&+\left. 6|x-y||x-z||z-t|\,\right) dxdydzdt+\mathcal{O}\left( \lambda
^{5}\right) .  \label{eq:Pert_exp_gen}
\end{eqnarray}
This expansion was derived from the zeros of a perturbative expansion for
the inverse of the Noyes form of the $T$ matrix. If we can calculate these
integrals exactly then we obtain the first terms of the perturbation
expansion for the dimensionless energy
\begin{equation}
\epsilon =\sum_{j=2}^{\infty }\epsilon ^{(j)}\lambda ^{j},
\label{eq:e_series}
\end{equation}
exactly.

Alternatively, one can derive this last expression directly by means of the
well known Rayleigh-Schr\"{o}dinger perturbation theory when the unperturbed
potential is $\beta v_{0}(x)=-\beta \delta (x)$, where $\delta (x)$ is the
Dirac delta. One thus obtains
\begin{equation}
\hat{\epsilon}=\sum_{j=2}^{\infty }\hat{\epsilon}^{(j)}(\beta )\lambda ^{j},
\label{eq:e_series_beta}
\end{equation}
and then recovers the actual $\lambda $-power series in the limit $\beta
\rightarrow 0$\cite{GR93}. The perturbation expansions (\ref{eq:Pert_exp_gen}%
) and (\ref{eq:e_series}) are valid for sufficiently small values of $%
\lambda $ and are based on the asymptotic boundary conditions $%
\lim\limits_{|x|\rightarrow \infty }\psi (x)=0$ for the eigenfunction $\psi
(x)$.

On the other hand, the perturbation expansion proposed by Lisowski et al\cite
{LNPSSG16} is based on periodic boundary conditions $\psi (x+L)=\psi (x)$.
In this case the eigenfunctions and eigenvalues of the unperturbed problem ($%
\lambda =0$) are
\begin{eqnarray}
\left| n\right\rangle &=&\frac{1}{\sqrt{L}}\exp \left[ \frac{2\pi inx}{L}%
\right] ,\;n=0,\pm 1,\pm 2,\ldots ,  \nonumber \\
e_{n} &=&\frac{4n^{2}\pi ^{2}}{L^{2}},  \label{eq:unpertubed_periodic}
\end{eqnarray}
and the coefficients of the perturbation expansion
\begin{equation}
\tilde{\epsilon}(L)=\sum_{j=2}^{\infty }\tilde{\epsilon}^{(j)}(L)\lambda
^{j},  \label{eq:e_series_perio}
\end{equation}
depend on the box length $L$. In principle, the exact solution of the
problem with periodic boundary conditions should approach the exact solution
of the problem with infinite boundary conditions in the limit $L\rightarrow
\infty $. One of the questions that we investigate in what follows is if it
is possible to recover the actual series (\ref{eq:e_series}) from the
approximate one (\ref{eq:e_series_perio}) when $L\rightarrow \infty $.

In order to facilitate the discussion in subsequent sections, from now on we
call these three approaches the $T$-method, the $\beta $-method and the $L$%
-method, respectively.

In addition to the series for small $\lambda $ we can also derive an
expansion for large $\lambda $ provided that we can expand $v(x)$ in a
Taylor series about $x=0$, $v(x)=v(0)+\frac{1}{2}v^{\prime \prime
}(0)x^{2}+\ldots $\cite{F01}. The first two terms are
\begin{equation}
\epsilon _{n}=-\lambda +(2n+1)\sqrt{\frac{\lambda v^{\prime \prime }(0)}{2}}%
+\ldots ,\;n=0,1,\ldots  \label{eq:e_series_large_lambda}
\end{equation}

\section{Examples}

\label{sec:examples}

In this section we analyze the perturbation series outlined above by means
of some exactly solvable examples. The first one is given by $v(x)=-1/\cosh
^{2}x$ that supports the bound-state energies\cite{F99}
\begin{equation}
\epsilon _{n}=-\left( \xi -n-1\right) ^{2},\;\xi =\frac{1+\sqrt{1+4\lambda }%
}{2},  \label{eq:e_n_Poschl}
\end{equation}
where the quantum number is restricted to $n=0,1,\ldots ,\xi -1$. The
perturbation series for all the eigenvalues have the same radius of
convergence $R=1/4$ that is determined by the branch-point singularity at $%
\lambda _{c}=-1/4$. For example, for the ground state we have
\begin{equation}
\epsilon _{0}=\frac{\sqrt{1+4\lambda }-1-2\lambda }{2}=-\lambda
^{2}+2\lambda ^{3}-5\lambda ^{4}+14\lambda ^{5}+\ldots .
\label{eq:e_0_Poschl}
\end{equation}
Contrary to what Lisowski et al suggested, here we have a perturbation
series for the eigenvalues of the quantum-mechanical problem valid at least
for $\lambda <1/4$. For larger values of the strength parameter we can
resort to any suitable summation method, for example, Pad\'{e} approximants%
\cite{BO78,BG96} and, even better, quadratic Pad\'{e} approximants\cite{BG96}%
. Note that in the present case the quadratic Pad\'{e} approximant $%
w^{2}+w\left( 2\lambda +1\right) +\lambda ^{2}=0$ yields the exact result $%
w(\lambda )=\epsilon _{0}(\lambda )$ for all values of $\lambda $. We can
also build two-point Pad\'{e} approximants\cite{BG96} that match the small-$%
\lambda $ and large-$\lambda $ series mentioned in section~\ref
{sec:gen_wells}.

The next example is given by the square potential
\begin{equation}
v(x)=\left\{
\begin{array}{c}
-1\;\mathrm{if\;}|x|\leq 1 \\
0\;\mathrm{if\;}|x|>1
\end{array}
\right. .
\end{equation}
The eigenvalues with even-parity eigenfunctions are solutions to the
implicit equation
\begin{equation}
k_{1}\tan k_{1}-k=0,\;k=\sqrt{-\epsilon },\;k_{1}=\sqrt{\epsilon +\lambda },
\label{eq:cuant_cond_sqr_pot}
\end{equation}
from which we derive the perturbation expansion
\begin{equation}
\epsilon =-\lambda ^{2}+\frac{4\lambda ^{3}}{3}-\frac{92\lambda ^{4}}{45}+%
\frac{1072\lambda ^{5}}{315}-\frac{84752}{14175}\lambda ^{6}+\ldots ,
\label{eq:e_series_square_pot}
\end{equation}
for the ground state. It agrees with the general expression (\ref
{eq:Pert_exp_gen}) through the fourth term.

In order to obtain the branch points in this case we take into account that
if $\left( d\lambda /d\epsilon \right) (\epsilon =\epsilon _{c})=0$ and $%
\left( d^{2}\lambda /d^{2}\epsilon \right) (\epsilon =\epsilon _{c})\neq 0$
then $\lambda \approx \lambda _{c}+A\left( \epsilon -\epsilon _{c}\right)
^{2}$, where $\lambda _{c}=\lambda \left( \epsilon _{c}\right) $, and $%
\epsilon \approx \epsilon _{c}+A^{-1/2}\sqrt{\lambda -\lambda _{c}}$. If the
eigenvalues are determined by an implicit equation of the form $F(\epsilon
,\lambda )=0$, then it follows from $dF/d\epsilon =(\partial F/\partial
\lambda )d\lambda /d\epsilon +\partial F/\partial \epsilon =0$ that $\lambda
_{c}$ and $\epsilon _{c}$ are determined by the pair of equations $%
F(\epsilon ,\lambda )=0$, $\partial F(\epsilon ,\lambda )/\partial \epsilon
=0$. In this way we obtain $\lambda _{c}=-0.4392288398$ and $\epsilon
_{c}=-1 $. As in the preceding case Pad\'{e} and quadratic Pad\'{e}
approximants yield accurate results for the ground-state eigenvalue in a
wide range of values of the strength parameter. In particular two-point
Pad\'{e} approximants (even of low order) yield considerably accurate
results for all values of $\lambda $. This potential cannot be expanded in a
Taylor series about origin but the ground-state eigenvalue behaves
asymptotically as $\epsilon =-\lambda +\mathcal{O}(1)$ for large $\lambda $.

This model is suitable for illustrating the application of the $\beta $%
-method of Gat and Rosenstein\cite{GR93}. To this end we solve the
Schr\"{o}dinger equation with the potential $\lambda v(x)-\beta \delta (x)$
and obtain the following quantization condition for the eigenvalues with
eigenfunctions of even parity:
\begin{equation}
k=\frac{k_{1}\left[ \beta \cos {\left( k_{1}\right) }+2k_{1}\sin {\left(
k_{1}\right) }\right] }{2k_{1}\cos \left( k_{1}\right) -\beta \sin {\left(
k_{1}\right) }}.  \label{eq:cuant_cond_sqr_pot_beta}
\end{equation}
This expression becomes equation (\ref{eq:cuant_cond_sqr_pot}) when $\beta
=0 $ as expected. If we expand $\epsilon $ in a Taylor series about $\lambda
=0$ we obtain the first terms of the expansion (\ref{eq:e_series_beta}):
\begin{eqnarray}
\hat{\epsilon}^{(0)}(\beta ) &=&-\frac{\beta ^{2}}{4},  \nonumber \\
\hat{\epsilon}^{(1)}(\beta ) &=&e^{-\beta }-1,  \nonumber \\
\hat{\epsilon}^{(2)}(\beta ) &=&\frac{2e^{-2\beta }\left( 1+\beta -e^{\beta
}\right) }{\beta ^{2}},  \nonumber \\
\hat{\epsilon}^{(3)}(\beta ) &=&\frac{e^{-3\beta }\left( 5e^{2\beta
}-8e^{\beta }\left( \beta +2\right) +6\beta ^{2}+14\beta +11\right) }{\beta
^{4}}.
\end{eqnarray}
What is important here is that these perturbation coefficients tend to those
in the expansion (\ref{eq:e_series_square_pot}) when $\beta \rightarrow 0$.

The next example is the Dirac-delta-potential $v(x)=-\delta (x)$ already
studied by Lisowski et al\cite{LNPSSG16}. Upon choosing the exact boundary
conditions $\lim\limits_{|x|\rightarrow \infty }\psi (x)=0$ we obtain the
dimensionless energy $\epsilon =-\lambda ^{2}/4$ for the only bound-state
eigenvalue. Note that both the $T$-method and the $\beta $-method discussed
in section~\ref{sec:gen_wells} yield the exact result. On the other hand, if
we solve the Schr\"{o}dinger equation with the periodic boundary conditions $%
\psi \left( -L/2\right) =\psi \left( L/2\right) $ and $\psi ^{\prime }\left(
-L/2\right) =\psi ^{\prime }\left( L/2\right) $ the eigenvalue is a root of
\begin{equation}
e^{-kL}\left( 2k+\lambda \right) -2k+\lambda =0,\;k=\sqrt{-\epsilon }.
\label{eq:quant_cond_delta}
\end{equation}
From this expression we obtain
\begin{equation}
\epsilon =-\frac{\lambda ^{2}}{4}\frac{\left( 1+e^{-kL}\right) ^{2}}{\left(
1-e^{-kL}\right) ^{2}}=-\frac{\lambda ^{2}}{4}\left(
1+4e^{-kL}+8e^{-2kL}+12e^{-3kL}+\ldots \right) ,
\end{equation}
which clearly shows that the error with respect to the exact result is of
the order of $e^{-kL}$ and, since $k$ decreases with $\lambda $, we
appreciate the necessity of increasing $L$ as $\lambda $ decreases as the
authors concluded from numerical analysis.

If we expand the solution to equation (\ref{eq:quant_cond_delta}) in a
Taylor series about $\lambda =0$ we obtain
\begin{equation}
\epsilon =-\frac{\lambda }{L}-\frac{\lambda ^{2}}{12}-\frac{L\lambda ^{3}}{%
180}-\frac{L^{2}\lambda ^{4}}{3780}-\frac{L^{3}\lambda ^{5}}{226800}+\ldots
\label{eq:e_series_Dirac_per}
\end{equation}
that depends on both $\lambda $ and $L$. The coefficients of this expansion
can also be obtained by means the well known Rayleigh-Schr\"{o}dinger
perturbation theory; for example:
\begin{eqnarray}
\tilde{\epsilon}^{(1)} &=&\left\langle 0\right| v\left| 0\right\rangle =-%
\frac{1}{L},  \nonumber \\
\tilde{\epsilon}^{(2)} &=&\sum_{n\neq 0}\frac{\left| \left\langle n\right|
v\left| 0\right\rangle \right| ^{2}}{e_{0}-e_{n}}=-\frac{1}{2\pi ^{2}}%
\sum_{n=1}^{\infty }\frac{1}{n^{2}}=-\frac{1}{12}.
\end{eqnarray}
While the $T$-method and $\beta $-method yield the exact result, the $L$%
-method gives rise to an infinite series. Furthermore, although equation (%
\ref{eq:quant_cond_delta}) yields the exact result when $L\rightarrow \infty
$ for fixed $\lambda $, the coefficients of the series (\ref
{eq:e_series_Dirac_per}) blow up. We conclude that the $L$-method
perturbation series is not asymptotic to the eigenvalue of the actual
quantum-mechanical problem but to the eigenvalue of the secular equation $%
\mathbf{HC}=\epsilon \mathbf{C}$, where $\mathbf{H}$ is the matrix
representation of the Hamiltonian operator in the basis set of periodic
functions (\ref{eq:unpertubed_periodic}) and $\mathbf{C}$ is a column vector
with the expansion coefficients for the wavefunction. The eigenvalue of this
matrix equation may be reasonably accurate provided that $L$ is suitably
chosen and the number of basis functions is sufficiently large. Besides, the
approximate $L$-method perturbation series (i.e. obtained with a finite
basis set) should approach the exact one (\ref{eq:e_series_Dirac_per}) as
the number of basis functions increases. What is clear from the results
above is that the $L$-method perturbation series may bear no resemblance
with the actual $\lambda $-series expansion given by either the $T$-method
or the $\beta $-method discussed in section~\ref{sec:gen_wells}. Both the $%
\beta $-method and the $L$-method resort to auxiliary parameters ($\beta $
and $L$, respectively). However, while $\lim\limits_{\beta \rightarrow 0}%
\hat{\epsilon}^{(j)}(\beta )=\epsilon ^{(j)}$, $\lim\limits_{L\rightarrow
\infty }\tilde{\epsilon}^{(j)}(L)$ blows up.

It is worth noting that if instead of periodic boundary conditions we impose
Neumann ones $\psi ^{\prime }\left( \pm L/2\right) =0$ then we obtain
exactly the same quantization condition (\ref{eq:quant_cond_delta}).

Finally we consider the exponential potential $V(x)=-V_{0}e^{-b|x|}$ also
discussed by Lisowski et al\cite{LNPSSG16}. Upon choosing the length $\gamma
=1/b$ we obtain $v(x)=-e^{|x|}$, $\lambda =2mV_{0}/(\hbar ^{2}b^{2})$ and $%
\epsilon =2mE/(\hbar ^{2}b^{2})$. The solution to the Schr\"{o}dinger
equation can be expressed in terms of the Bessel function of the first kind $%
\psi (x)=AJ_{\nu }(z)$, where $A$ is a normalization constant, $\nu =2\sqrt{%
-\epsilon }$ and $z=2\sqrt{\lambda }e^{-x/2}$. The boundary condition for
even states at origin $\psi ^{\prime }(0)=0$ leads to
\begin{equation}
z_{0}J_{\nu +1}\left( z_{0}\right) -\nu J_{\nu }\left( z_{0}\right)
=0,\;z_{0}=2\sqrt{\lambda }.
\end{equation}
From this equation we obtain the following expansion for the dimensionless
ground-state energy
\begin{equation}
\epsilon =-\lambda ^{2}+3\lambda ^{3}-\frac{143\lambda ^{4}}{12}+\frac{%
3887\lambda ^{5}}{72}-\frac{71303\lambda ^{6}}{270}+\ldots ,
\label{eq:e_series_expo_pot}
\end{equation}
that agrees with the general expression (\ref{eq:Pert_exp_gen}) through the
fourth term. If we only keep the first term in the right-hand side we obtain
$E\approx -2mV_{0}^{2}/(\hbar ^{2}b^{2})$ that agrees with the result of
Lisowski et al\cite{LNPSSG16} when $\hbar =1$.

In this case we can also obtain reasonable results from the perturbation
series by means of Pad\'{e} approximants and quadratic Pad\'{e}
approximants. For example, the Pad\'{e} approximant $[3,3](\lambda )$
constructed from the expansion (\ref{eq:e_series_expo_pot}) yields
acceptable results for $0\leq \lambda <1$.

\section{Conclusions}

\label{sec:conclusions}

It is known since long ago that one can apply perturbation theory to the
Schr\"{o}dinger equation with a short-range potential $\lambda v(x)$ and
obtain a suitable $\lambda $-power series asymptotic to the ground-state
eigenvalue\cite{P80} (and references therein). In section~\ref{sec:gen_wells}
we mentioned two approaches for that purpose. Lisowski et al\cite{LNPSSG16}
proposed the $L$-method for the construction of a perturbation series
starting from an unperturbed model with periodic boundary conditions.
Although the exact eigenvalue of the Schr\"{o}dinger equation with periodic
boundary conditions tends to the eigenvalue of the Schr\"{o}dinger equation
with boundary conditions at infinity as $L\rightarrow \infty $ the same does
not occur in the case of the $L$-method perturbation series because its
coefficients blow up when $L\rightarrow \infty $. Therefore, the $L$-method
power series is asymptotic to the eigenvalue of the problem with periodic
boundary conditions for a given $L$ and never to the actual physical
eigenvalue. Since the authors resorted to a matrix representation of the
Hamiltonian operator they did not even obtain the exact $L$-method
perturbation series because of a necessary truncation of the basis set.
Their perturbation series is asymptotic to the eigenvalue of the matrix
representation of the Hamiltonian operator. If $L$ is sufficiently small the
series exhibits good convergence properties but the result is far from the
eigenvalue of the problem with infinite boundary conditions. On the other
hand, if $L$ is sufficiently large the convergence properties are poor and
one is forced to resort to an efficient summation method. It seems by far
more convenient to diagonalize the matrix representation of the Hamiltonian
for a sufficiently large value of the box length $L$ and a sufficiently
large number of basis functions.

\end{document}